\documentclass{vgtc}                          

\graphicspath{{figures/}{pictures/}{images/}{./}} 

\usepackage{times}                     

\usepackage{tabu}                      
\usepackage{booktabs}                  
\usepackage{lipsum}                    
\usepackage{mwe}                       

\usepackage{mathptmx}                  

\onlineid{0}

\vgtccategory{Research}

\vgtcinsertpkg

\title{Integrated Telehealth and Extended Reality to Enhance Home
Exercise Adherence Following Total Hip and Knee Arthroplasty}

\author{Christy L. Conroy\thanks{e-mail: christy.conroy@brooksrehab.org}\\ %
        \scriptsize Brooks Rehabilitation %
        \and
    Gina M. Brunetti\thanks{e-mail: gina.brunetti@brooksrehab.org}\\ %
        \scriptsize Brooks Rehabilitation %
    \and
    Angelos Barmpoutis\thanks{e-mail: angelos@digitalworlds.ufl.edu}\\ %
        \scriptsize University of Florida %
    \and
    Emily J. Fox\thanks{e-mail: ejfox@phhp.ufl.edu}\\ %
        \parbox{1.4in}{\scriptsize \centering University of Florida \\ Brooks Rehabilitation} %
        }

\teaser{
  \centering
  \includegraphics[width=1.0\linewidth]{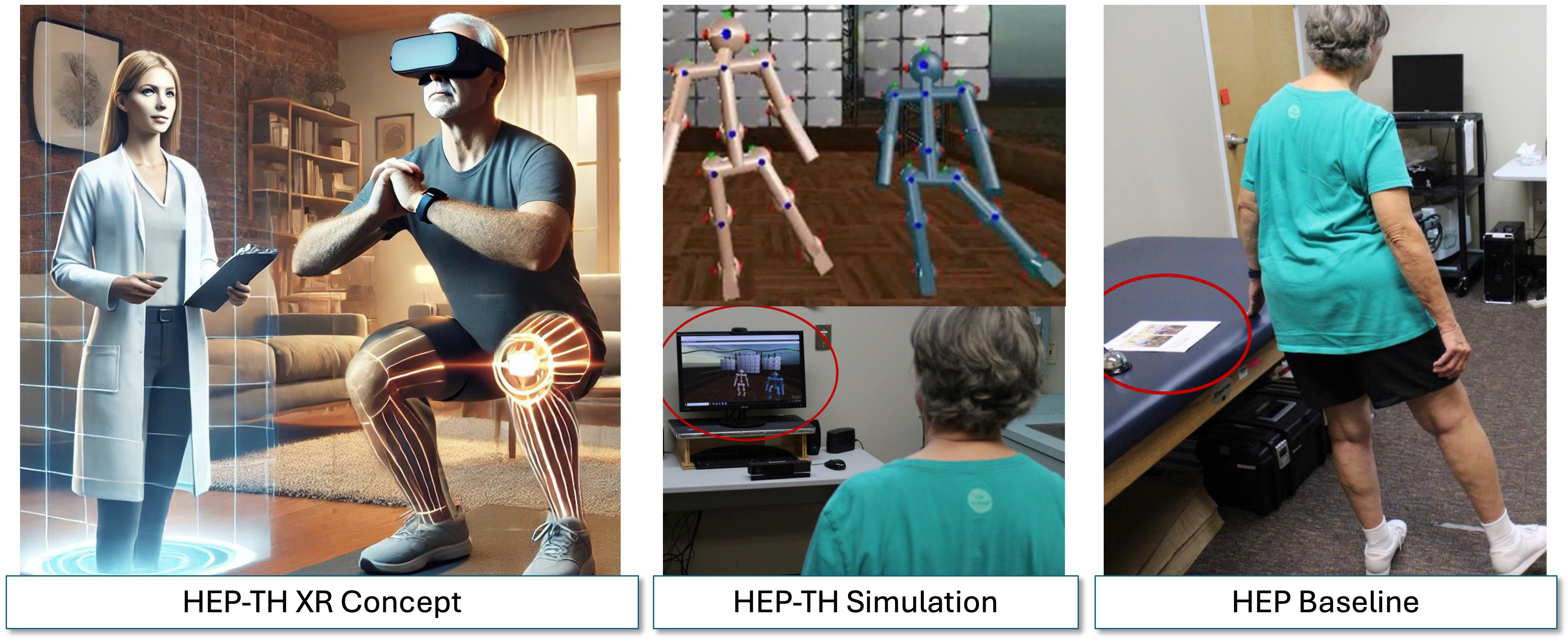}
  \caption{Left: Illustration of the proposed XR-based concept that combines home exercise programs (HEP) and telehealth (TH). Center: Exercise with the developed Tele-PhyT system (experimental condition). Right: Independent Exercise (control condition). }
  \label{fig:teaser}
}

\abstract{
Nearly one million total hip and knee arthroplasties (THA/TKA) are performed annually in the United States, with most patients discharged home and prescribed home exercise programs (HEPs) to enhance lower extremity function. Traditional paper-based HEPs, while accessible and low-cost, often lack engagement and real-time feedback, which are critical for adherence and performance optimization. Extended reality (XR) and telehealth (TH) systems offer promising solutions, combining engagement and feedback, though each has limitations. To address these gaps, we designed and executed a pilot study that compared exercise performance in individuals with THA/TKA using a conventional paper-based HEP versus a proof-of-concept system, dubbed Tele-PhyT, that included the ideal characteristics of a future XR technology that would enable seamless HEP-TH systems, with robust marker-less full body tracking, real-time visual feedback, and performance quantification. The pilot study used a randomized cross-over design and targeted two types of users: therapists and patients. Participants favored Tele-PhyT for its real-time feedback and ease of use, and noted its potential to improve HEP adherence and exercise accuracy.
} 


\begin{document}


\firstsection{Introduction}
\maketitle

Nearly one million total hip and knee arthroplasties (THA/TKA) are performed annually in the
United States \cite{bib1}. In 74\%-96\% of cases, patients are discharged directly home versus other care settings
\cite{bib2,bib3} and home exercise programs (HEPs) are commonly prescribed to increase lower extremity range of
motion, strength, and functional mobility \cite{bib4,bib5,bib6}. While conventional paper-based HEPs are simple, lowcost,
and accessible, a major limitation is lack of patient engagement and feedback on exercise
performance \cite{bib7}. These are critical issues since engagement increases motivation and exercise adherence
\cite{bib8} and real-time feedback promotes correct movement patterns and encourages maximal
performance \cite{bib9,bib10,bib11}. Therefore, strategies to improve engagement and provide real-time feedback
associated with HEPs could promote exercise adherence and improve outcomes following knee and hip
joint arthroplasty \cite{jcm11071766}.

Extended reality (XR) and telehealth (TH) systems are potential cost-effective strategies to
promote patient engagement and enable HEP performance feedback \cite{bib10,bib12,bib13}. The use of these
systems for remote rehabilitation services during the global Covid-19 pandemic in the United States
increased more than 50-154\% (depending on the time period) and ongoing use of various technologies
is anticipated \cite{bib14,bib15}. While utilization of XR and TH systems has gained popularity, each strategy has
advantages and limitations \cite{bib16,bib17}. TH systems, defined here as web-based audiovisual interactions,
often promote real-time (synchronous) patient-therapist interactions, but typically do not allow real-time
monitoring or quantification of exercise performance and feedback, and often lack motivational
features. Conversely, asynchronous (not real-time) virtual or augmented reality systems utilizing virtual agents and immersive
experiences are well established for promoting engagement and motivation through gaming features
and visual feedback; however, these systems often do not support real-time patient-therapist
interactions and performance feedback related to quality of movement \cite{bib18,bib19,bib20}. Thus, given the pros and
cons of these systems, it would be valuable to integrate XR and TH features into an intuitive and, easy to use system.

Our group identified the key technological requirements for such an XR concept design, which could include a lightweight wearable headset with a wide field of view, robust full-body tracking, real-time streaming and rendering of participants in the same 3D space, audiovisual feedback, and features to quantify exercise performance, as illustrated in the conceptual design shown in Fig.\ref{fig:teaser} (left). Based on the current market analysis of consumer XR devices (Meta Quest Pro, Apple Vision Pro, HoloLens 2, and Magic Leap 2), significant progress has been made toward making these devices more affordable and capable of tracking hand movements. However, there is still room for improvement in full-body tracking, field of view, and size/weight of headsets. Although these limitations make current devices unfit and/or unsafe for clinical use, this progress suggests that within the next decade such devices could become available to the consumer market. 

Patients with hip and knee arthroplasties might benefit from use of such a system to support HEP performance. Investigations of XR and TH use for exercise motivation and performance in this population is particularly limited and few low cost systems for home rehabilitation use are available \cite{bib21,bib22,jcm11071766}. Therefore, designing and studying a HEP-TH proof-of-concept that meets the functional specifications of such system could provide valuable insights for building future XR-based strategies that promote exercise adherence and improve outcomes following knee and hip joint arthroplasty. Accordingly, our group developed a custom system that aims to simulate HEP-TH by incorporating many of these features into a clinically safe  experimental setup with a marker-less depth sensor and virtual reality rendering in a computer monitor. The system, referred to as Tele-PhyT allows for real-time body tracking and streaming, incorporates visual feedback regarding exercise performance, and includes features to quantify exercise performance. 

Our objective was to evaluate the potential value of HEP-TH systems for performance of a HEP. The primary aim of this proof-of-concept study was to compare lower extremity exercise performance in individuals with hip and knee total joint arthroplasties using a conventional HEP (with paper-based pictures and instructions) versus the custom Tele-PhyT. Our secondary aim was to obtain qualitative feedback from patient and therapist participants regarding the perceived potential usability of the Tele-PhyT system, its value for exercise engagement.

The key contributions of this work are threefold: 1) we developed a clinically safe simulation environment that implements the key functional characteristics of our HEP-TH XR concept, 2) we enrolled real patients with hip or knee arthroplasty, and physical or occupational therapists to conduct our pilot study, and 3) we discussed the results obtained from quantitative and qualitative data.

\section{Methods}

\subsection{Design Features, Capabilities, and Equipment}
The proof-of-concept Tele-PhyT system captures and streams 3D body motion in real time and is used to animate human avatars of remote users (e.g., patients and therapists) within the same 3D space projected in a 2D display. The system utilizes our custom internet streaming software developed in JavaScript using WebSockets and WebGL and an infrared depth camera (Microsoft Kinect 2), which provides robust, markerless full-body tracking using the J4K library that it was first introduced in \cite{digitalWorlds:122}. This experimental setup is clinically safe to be used in our study and meets the specifications envisioned for an ideal future XR system, as described in Sec.1. 

This system enables natural human-computer interaction using simple body gestures. Remote users can interact naturally with each other through body gestures or voice, combining the benefits of XR and TH. The low bandwidth requirements of the recorded motion data and audio allow the system to operate over most home-based internet connections. Additionally, the real-time transmission of motion data enables graphical feedback for users. Quantities such as movement magnitudes and velocities can be estimated from the captured 3D data and recorded for therapists to monitor patient progress. This proof-of-concept design implements most of the specifications described in Sec.1 and allowed us to execute a future looking pilot study while avoiding safety or user bias issues caused by limitations of current XR systems, such as heavy or bulky headsets, partial or no body tracking, and other constraints.

\subsection{Participants}
Participants were recruited from a metropolitan rehabilitation health system. Inclusion criteria for
patient participants included community-dwelling adults 40 to 90 years of age who were $>2$ weeks post
lower extremity total joint arthroplasty. Individuals confirmed no uncorrected visual impairment, no
severe medical conditions or injuries limiting mobility (aside from the recent joint arthroplasty), and the
ability to walk independently, with a device if needed. Inclusion criteria for the therapist participants
included an active license as a Physical or Occupational Therapist or Therapist Assistant (PT, OT, COTA respectively in Table \ref{tab:demographics1}). The study was
approved by the University of Florida Institutional Review Board. All participants provided written
informed consent in accordance with the Declaration of Helsinki. Table \ref{tab:demographics1} reports the demographics of patients and therapists participated in this study. The table also reports the setting of therapists' clinical experience, which includes outpatient (O), inpatient (I), acute care (A), home health (H), skilled nursing facility (SNF), and school system (S).

\begin{table}[tb]
  \caption{Patient and therapist participant demographics.}
  \label{tab:demographics1}
  \scriptsize%
	\centering%
  \begin{tabu}{%
	p{1.0cm}%
	*{5}{c}%
	}
  \toprule
   Patient & Age & Sex & \rotatebox{90}{Months} \rotatebox{90}{since} \rotatebox{90}{injury}&    \rotatebox{90}{Type of} \rotatebox{90}{Surgery} &   Laterality    \\
  \midrule
    VRP-05 & 68 & Female  & 5 & TKA & Right \\
    VRP-06 & 64 & Female & 3 & TKA & Right \\
    VRP-07 & 68 & Female & 3 & TKA & Right \\
    VRP-08 & 76 & Male & 2 & THA & Right \\
    VRP-09 & 62 & Male & 5 & TKA & Left \\
    VRP-11 & 58 & Female & 9 & THA & Left \\
    VRP-12 & 60 & Female & 1 & TKA & Right \\
    VRP-13 & 49 & Male & 1 & TKA & Left \\
    VRP-15 & 51 & Female & 3 & TKA & Right \\
    VRP-16 & 68 & Female & 2 & THA & Right \\
    VRP-17 & 65 & Female & 2 & THA & Left \\
    VRP-18 & 61 & Female & 2 & TKA & Left \\
    VRP-19 & 72 & Female & 3 & THA & Left \\
     \midrule
   Average & 63.23 & 10 F & 3.15 & 8 TKA & 7 R\\
         & 63.23 & 3 M & 3.15 & 5 THA & 6 L \\
   \midrule
    SD & ±7.68 & & ±2.15 & & \\
    \midrule
     Therapist & Age & Sex & \rotatebox{90}{Years} \rotatebox{90}{practicing} &    Discipline &   \rotatebox{90}{Setting of} \rotatebox{90}{experience}    \\
  \midrule
  VRT-08 & 28 & Female & 3.5 &  PT & O,I \\
VRT-09 & 52 & Female & 22 & OT & O,I,A,H \\
VRT-10 & 33 & Female & 10 & OT & O \\
VRT-11 & 31 & Female & 5 & COTA & O,I,SNF\\
VRT-12 & 30 & Female & 5 & PT & O \\
VRT-13 & 28 & Female & 4 & PT & O,A,S\\
VRT-14 & 41 & Female & 16 & PT & O \\
VRT-15 & 39 & Female & 13 & PT & O \\
VRT-16 &  & Female & 2.5 & OT & \\
VRT-17 & 26 & Female & 2 & OT & O\\
VRT-18 & 28 & Female &  0.5 & PT & O,I,SNF \\
\midrule
Average & 33.6 & 11 F & 7.59 & 6 PT & \\ 
        &       & 0 M  & & 5 OT/COTA & \\
SD & ±8.13 & & ±6.49 & & \\
  \bottomrule
  \end{tabu}%
\end{table}

\subsection{Protocol}

\subsubsection{Study Design}
A single session, random order cross-over design was used to assess exercise performance under two
conditions—a control condition representing an independent HEP using paper-based instructions (Fig.\ref{fig:teaser} right), and
an experimental condition involving the Tele-PhyT system (Fig.\ref{fig:teaser} center). Sessions were led by a licensed
physical therapist familiar with the patient population (referred to as remote therapist). The same lower
extremity exercises were performed under both conditions. Therapist participants observed the Tele-PhyT session in a separate room. Basic demographic data were collected via questionnaires. Participants also reported familiarity with technology such as computers, home-based commercial video games, VR and TH use. Patient participants were oriented to four common lower
extremity standing exercises at the start of the session with basic instructions to avoid compensatory
trunk leaning, and how to perform a squat. These typical rehabilitation exercises included hip abduction
and hip extension with the knee extended, single-limb hip flexion with the knee flexed, and squats.
During each condition, exercises were performed in random order for 12-15 repetitions for each lower
extremity and always commenced with the right lower extremity. Participants were instructed to
stabilize themselves with upper extremity support using a chair or elevated mat table as needed to
prevent loss of balance. Exercise data were acquired using the infrared depth camera for both
conditions. Questionnaires were completed at the end of the session to solicit opinions regarding the Tele-PhyT system.

\subsubsection{Experimental Condition: Exercise with Tele-PhyT}
Use of the Tele-PhyT system involved exercising about 6 feet from a 24” computer screen and the infrared depth camera. The screen displayed the participant and remote therapist avatars within the same 3D space (Fig.\ref{fig:teaser} center).
The remote therapist, located in an adjacent room not known to the patient participant, provided
standardized verbal feedback midway through each exercise for posture and movement corrections,
similar to the initial instructions. 

\subsubsection{Control Condition: Independent Exercise}
The goal of the control condition was to replicate features of independent exercise using a written HEP.
The initial introduction to the exercises served as the therapist-guided review of the exercises listed on
the written HEP, which included a paper-based set of instructions with pictures (Fig.\ref{fig:teaser} right). The
participants performed the exercises alone in a room and were provided a call bell in case they needed
assistance. The participant stood in front of the powered-off computer screen and obscured infrared
depth camera which enabled remote monitoring and recording of the participants’ movement from the Tele-PhyT system without their awareness.

\subsubsection{Questionnaires}
Patient and therapist participants completed questionnaires at the end of the session. To minimize bias,
questionnaires were administered by a different study staff member from the remote therapist. The
questionnaires utilized both Likert scale and open-ended questions regarding:
1) Tele-PhyT usability which includes the perceived ease and convenience of using Tele-PhyT to
guide a HEP, and ease of interactions between the remote therapist and patient.
2) Perceived value of Tele-PhyT use which includes exercise engagement, adherence, and
preferences while using the Tele-PhyT system versus other exercise instruction methods.

\subsection{Data Acquisition and Analysis}
Data were acquired from the infrared depth camera and processed to calculate movement magnitudes and velocities. Eight to twelve repetitions of each exercise were analyzed.
Repetitions were excluded based on lack of clearly defined movement peaks (i.e. participant moved the
limb out of camera view). The kinematic data of both hip joints for each exercise were low pass filtered
(4 Hz), exported and plotted in Visual 3D software to enable inspection and visualization. The start and
end of each exercise repetition were manually labeled.
Exercise performance was analyzed in two ways: maximal joint movement and movement velocity for
each exercise repetition. The maximal joint movement was identified as the peak joint angle $\theta$
after the ‘start’ label for the selected joint during each exercise repetition. Movement velocity was
calculated based on: $(\theta_{peak}-\theta_0)/(t_{peak}-t_0)$, where $\theta_{peak}$ and $\theta_0$ denote the peak and starting joint angle respectively.
The values for each repetition were averaged for each lower extremity and exercise. Data for each
exercise were compared between the intact and impaired limbs, and between the experimental and
control conditions.
Questionnaire data from Likert Scale responses were compiled for reporting and open-ended responses were grouped to identify common
themes.

Data were assessed for normality using the Shapiro-Wilk test. Normally distributed data were assessed
using dependent t tests. Non-normally distributed data were assessed for differences using the
Wilcoxon signed-rank test. Descriptive and summary statistics are provided. Results are presented as
means ± standard deviation. Statistical significance was established at p<.05. Statistics were calculated
using IBM SPSS Statistics for Windows, version 25. Uncorrected p values
are reported \cite{bib23,bib24,bib25}.

\section{Results}
Thirteen adults (10 females, 63.23±7.68 years old) with hip or knee arthroplasty (8 TKA, 7 right-sided,
3.15±2.15 years since surgery) completed the study as patient participants. Eleven licensed physical or
occupational therapists or therapy assistants (all female, 6 physical therapists, 33.6±8.13 years old,
7.59±6.49 years practicing) completed the study as therapist participants.


\subsection{Exercise Performance}

\begin{figure}[tb]
 \centering 
 \includegraphics[width=1.0\columnwidth]{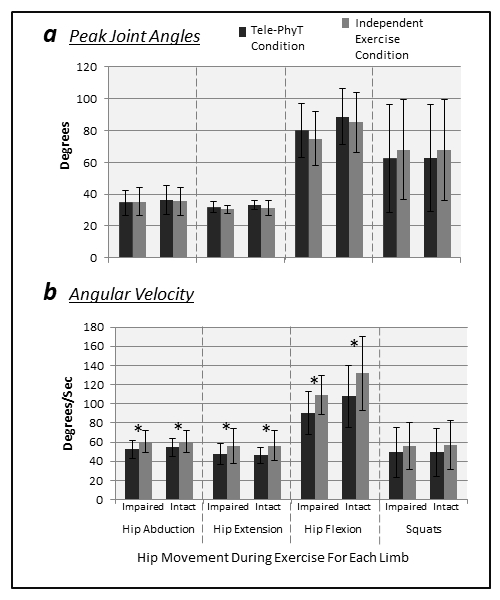}
 \caption{Comparison of the peak joint angles and angular velocities between the test and control conditions.}
 \label{fig:tracking_data}
\end{figure}

Movement velocities, were slower for all exercises except squats for both limbs during the
Tele-PhyT condition ($p<.05$ for hip abduction, extension, and flexion; $p=0.3$ for hip flexion during squats;
Fig.\ref{fig:tracking_data}b). During the Tele-PhyT condition (and Independent HEP condition), the mean velocity for each
exercise of the impaired and intact limb, respectively, was: impaired limb hip abduction 52.5±9.4°/s
(60.3±11.6°/s), intact limb 54.5±9.5°/s (60.5±11.7°/s); impaired limb hip extension 47.4±11.2°/s
(55.9±18.6°/s), intact limb 46.1±8.4°/s (56.2±15.7°/s); impaired limb hip flexion 90.2±22.2°/s
(108.8±20.4°/s), intact limb 107.7±32.5°/s (131.8±38.4°/s); and hip flexion during squats impaired limb
49.2±26.3°/s (56.0±24.2°/s), intact limb 49.5±25.2°/s (56.7±25.5°/s). The slower speed during the Tele-PhyT condition may have been influenced by the remote therapist’s exercise interactions. 

However, there were no significant differences in joint movement magnitudes between the two exercise conditions ($p>.05$; Fig.\ref{fig:tracking_data}a).

\subsection{Questionnaire Data}

Overall, patient and therapist participants agreed it was easy to communicate through the Tele-PhyT
system. Patients indicated it was easy to understand how to interact within the system and therapists
felt the system was effective for hosting a real-time session. Majority of patients reported they
preferred the use of the Tele-PhyT system to complete a HEP over doing exercises on their own via
traditional HEP (Table \ref{tab:vis_papers}).

Nearly all participants agreed that it was easy to understand and implement remote therapist feedback
and adjust exercise performance during the Tele-PhyT condition. One patient participant commented
that her favorite aspect was “making sure that I am performing exercises correctly”. Eight of eleven
therapist participants and seven of thirteen patient participants indicated the aspect they liked most
was real-time feedback to correct posture and/or improve exercise performance.

Over half of all patient and therapist participants believed Tele-PhyT would increase patient adherence
with prescribed HEPs. Open-ended comments from participants indicated the Tele-PhyT system was
“entertaining”, was “more engaging working with the [remote] therapist”, and “more engaging than a written HEP”.

Over half of all patient and therapist participants were interested in incorporating the use of Tele-PhyT
into current practice. All therapist participants and over half of all patient participants preferred
exercise with Tele-PhyT versus a conventional gaming system (i.e., Wii) that does not include live
interactions with a remote therapist.

Nearly all participants found the avatars and real-time verbal feedback provided by the remote therapist
helpful and over 70\% reported the graphical feedback was helpful. Patients reported it was “fun
watching [their own] avatar” and liked the “ability to correct [their own] posture” during the exercise
session, while therapists appreciated having “the therapist avatar as well for demonstration purposes
and not just verbal feedback”. One therapist participant indicated the graphical feedback “was
competitive for the patient by seeing the lines, to show the magnitude of movements”, while one
patient participant indicated it was “helpful to compare [my] impaired knee to [my] intact knee”.

\begin{table}[tb]
  \caption{User feedback on perceived ease of use and usefulness}
  \label{tab:vis_papers}
  \scriptsize%
	\centering%
  \begin{tabu}{%
	p{3.9cm}%
	*{5}{c}%
	}
  \toprule
   Patients' Questions & \rotatebox{90}{Disagree} &   \rotatebox{90}{Somewhat} \rotatebox{90}{Disagree}&   \rotatebox{90}{Neutral} &   \rotatebox{90}{Somewhat} \rotatebox{90}{Agree} &   \rotatebox{90}{Agree}    \\
  \midrule
    Ease of communication & 0 & 0  & 0 & 0 & 13 \\
    Understood \& used ther. feedback & 0 & 0 & 0 & 0 & 13\\
    Clarity of exercises & 0 & 0  & 0 & 0 & 13 \\
    Clarity of Tele-PhyT system & 0 & 0 & 0 & 0 & 13 \\
    Improved HEP workout & 0 &  1 & 4 & 0 & 8 \\
    Increased HEP compliance & 0 &  1 & 2 & 3 & 7 \\
    Likelihood to use Tele-PhyT & 1 & 0 & 1 & 2 & 9 \\
    \midrule
    Therapists' Questions & & & & \\
     \midrule
    Ease of communication & 0 & 0  & 0 & 1 & 10 \\
    Patients understood/used feedback & 1 & 0 & 0 & 1 & 9 \\
    Effectiveness of real-time session & 0 & 0  & 0 & 1 & 10 \\
    Ease of hosting Tele-PhyT session & 0 & 0  & 0 & 1 & 10 \\
    Improved HEP performance & 0 & 0  & 0 & 1 & 10 \\
    Increased HEP compliance & 0 & 0  & 1 & 3 & 7 \\
    Interest in adopting Tele-PhyT & 0 & 0  & 0 & 4 & 7 \\
  \bottomrule
  \end{tabu}%
\end{table}

\section{Discussion}
Study outcomes suggest adults post lower extremity joint replacement demonstrate different angular velocities but similar movement
magnitudes when performing exercises with the custom system, as
when exercising independently using a conventional HEP. However, the Tele-PhyT system enabled
remote, real-time interactions in a virtual environment allowing therapist instruction and feedback,
which participants reported to be valuable. Therapist participant comments supported potential use of
the system for home exercise sessions. Most notable, users reported preference for the Tele-PhyT
system to promote engagement and home exercise adherence. 
Overall, participant feedback supports integrated XR and TH features and their use to increase exercise adherence and allow for remote
monitoring. Further, an XR/TH system, can be adapted to allow for group exercise
sessions and recording of exercise data to monitor progress and patient care needs.

Overall, the Tele-PhyT allowed for varying types of feedback (visual, quantitative, verbal) regarding
knowledge of exercise performance which could be further adapted or customized to facilitate motor
learning. The system also could provide knowledge of results feedback within a single exercise session or
across multiple sessions. While XR gaming systems provide feedback on performance and game related
outcomes, these systems generally are not focused on rehabilitation or motor learning. Additionally,
these systems do not include real-time interactions with a therapist knowledgeable in musculoskeletal
conditions. Although patient participants reported positive opinions regarding all forms of feedback, the
verbal feedback from the remote therapist was rated the highest.

The positive opinions expressed by patient participants regarding the real-time therapist verbal
feedback is a potentially important outcome, and it is directly related to the ability of the remote therapist to assess in real time the streamed 3D motion data of the patients. Social interactions and/or having an exercise coach can
increase motivation and exercise adherence\cite{bib7,bib13,bib26,bib27,bib28}. Specifically, feedback that provides positive
reinforcement, assurance that exercises are performed correctly, and individualized coaching leads to
greater exercise adherence\cite{bib7}. Social interactions were particularly limited during the Covid-19
pandemic and tele-rehabilitation use increased substantially, demonstrating the importance of
maintaining social connections as well as the need for continued support and coaching during
rehabilitation even during times of quarantine\cite{bib7,bib26,bib27}.

In addition to patient-therapist interactions, the Tele-PhyT system provides opportunity for group
interactions to be guided by the remote therapist. Group exercise interactions are well established as
effective for rehabilitation of various musculoskeletal conditions \cite{bib29,bib30,bib31} and could increase social
interactions as well as the cost-effectiveness of care delivery\cite{bib32,bib33,bib34,bib35}. Peer interactions, encouragement,
and friendly competitions could be incorporated and are effective aspects of group exercise
sessions\cite{bib33,bib34}. Furthermore, these aspects may be particularly important for individuals who lack
family or caregiver support, or who are struggling with social isolation and depression\cite{bib33}. These
challenges impact as much as 40\% of the older adult population and are associated with higher care
costs and greater disability\cite{bib36,bib37}. While these aspects were not investigated in this study, these initial
study outcomes suggest further development of XR/TH systems. Specifically, development of dynamic systems that not only include virtually engaging environments, but also allow for real-time interactions with rehabilitation experts should be emphasized, particularly for those who may need to exercise at home but struggle to be compliant with prescribed home programs.

\section{Limitations}
This study involved a small number of participants and a higher proportion of individuals post TKA vs.
THA. However, our study population is representative of the general population where individuals
undergoing knee arthroplasty are nearly double those who undergo hip arthroplasty.[1] The study
design involved a single-day, mock home exercise session, which allowed for a proof-of-concept study to
gain understanding of this system and its features. This study design likely influenced outcomes which
may not represent exercise outcomes in the home. This mock arrangement ensured proper system
functioning including internet connections, while avoiding variable home environments that could
interfere with addressing the primary study questions. The study also involved a limited set of rehabilitation exercises. Future steps should explore a wider range of exercises in a home setting and/or across several sessions.

\section{Conclusion}
Exercise performance with use of the Tele-PhyT system is similar to performance during a self-guided
home exercise program. However, the Tele-PhyT system has additional beneficial features including the
ability to quantify exercise performance, track progress and interact real-time with a healthcare
provider. The majority of therapist and patient participants expressed that this system was overall easy
to use, would improve exercise performance, and would improve adherence with prescribed exercise
programs.

\section*{Figure Credits}
\label{sec:figure_credits}
The left panel of \Cref{fig:teaser} was generated by OpenAI DALL-E.

\acknowledgments{
The authors thank all participants for their contribution to this project. The authors thank Brooks Rehabilitation therapists, Brooks Clinical Research Center and Motion Analysis Center (Brooks Rehabilitation, Jacksonville, FL), especially Paul Freeborn for contributions to this project.

This project was supported by the University of Florida Informatics Institute Seed Fund Program, the Brooks-PHHP Research Collaboration, and the NIH K12 Grant [HDO55929, to EF].}

\bibliographystyle{abbrv-doi}

\bibliography{template}

\begin{thebibliography}{10}

\bibitem{bib26}
R.~Argent, A.~Daly, and B.~Caulfield.
\newblock Patient involvement with home-based exercise programs: Can connected health interventions influence adherence?
\newblock {\em JMIR MHealth and UHealth}, 6(3):e47, Mar. 2018.

\bibitem{bib4}
R.~Argent, P.~Slevin, A.~Bevilacqua, M.~Neligan, A.~Daly, and B.~Caulfield.
\newblock Clinician perceptions of a prototype wearable exercise biofeedback system for orthopaedic rehabilitation: A qualitative exploration.
\newblock {\em BMJ Open}, 8(10):e026326, Oct. 2018.

\bibitem{bib5}
R.~Argent, P.~Slevin, A.~Bevilacqua, M.~Neligan, A.~Daly, and B.~Caulfield.
\newblock Wearable sensor-based exercise biofeedback for orthopaedic rehabilitation: A mixed methods user evaluation of a prototype system.
\newblock {\em Sensors}, 19(2):N/A, Jan. 2019.

\bibitem{digitalWorlds:122}
A.~Barmpoutis.
\newblock Tensor body: Real-time reconstruction of the human body and avatar synthesis from rgb-d.
\newblock {\em IEEE Transactions on Cybernetics, Special issue on Computer Vision for RGB-D Sensors: Kinect and Its Applications}, 43(5):1347--1356, October 2013.

\bibitem{bib21}
A.~Barmpoutis, J.~Alzate, S.~Beekhuizen, H.~Delgado, P.~Donaldson, A.~Hall, et~al.
\newblock Assessment of haptic interaction for home-based physical tele-therapy using wearable devices and depth sensors.
\newblock {\em Studies in Health Technology and Informatics}, 220:33--38, 2016.

\bibitem{bib14}
O.~Bestsennyy, G.~Gilbert, A.~Harris, and J.~Rost.
\newblock Telehealth: A quarter-trillion-dollar post-covid-19 reality?
\newblock {\em McKinsey \& Company}, May 2020.

\bibitem{bib32}
C.~L. Coulter, J.~M. Weber, and J.~M. Scarvell.
\newblock Group physiotherapy provides similar outcomes for participants after joint replacement surgery as 1-to-1 physiotherapy: A sequential cohort study.
\newblock {\em Archives of Physical Medicine and Rehabilitation}, 90(10):1727--1733, Oct. 2009.

\bibitem{bib33}
N.~Dnes, B.~Coley, K.~Frisby, A.~Keller, J.~Suyom, C.~Tsui, et~al.
\newblock "a little bit of a guidance and a little bit of group support": A qualitative study of preferences, barriers, and facilitators to participating in community-based exercise opportunities among adults living with chronic pain.
\newblock {\em Disability and Rehabilitation}, 0(0):1--10, Mar. 2020.

\bibitem{bib27}
R.~Essery, A.~W.~A. Geraghty, S.~Kirby, and L.~Yardley.
\newblock Predictors of adherence to home-based physical therapies: A systematic review.
\newblock {\em Disability and Rehabilitation}, 39(6):519--534, Mar. 2017.

\bibitem{jcm11071766}
E.~Fascio, J.~A. Vitale, P.~Sirtori, G.~Peretti, G.~Banfi, and L.~Mangiavini.
\newblock Early virtual-reality-based home rehabilitation after total hip arthroplasty: A randomized controlled trial.
\newblock {\em Journal of Clinical Medicine}, 11(7), 2022. doi: {{%
10\hspace{.1pt}\discretionary{.}{%
}{.}\hspace{.4pt}3390\discretionary{/}{%
}{/}jcm11071766}}


\bibitem{bib22}
C.~S. Fernandes, B.~Magalhães, F.~Goncalves, P.~C. Nogueira, and C.~Santos.
\newblock The use of gamification in patients undergoing hip arthroplasty: Scoping review.
\newblock {\em Games for Health Journal}, May 2021.
\newblock [cited 2021 May 11].

\bibitem{bib2}
A.~N. Fleischman, M.~S. Austin, J.~J. Purtill, J.~Parvizi, and W.~J. Hozack.
\newblock Patients living alone can be safely discharged directly home after total joint arthroplasty: A prospective cohort study.
\newblock {\em The Journal of Bone and Joint Surgery. Am.}, 100(2):99--106, Jan. 2018.

\bibitem{bib9}
M.~Friedrich, T.~Cermak, and P.~Maderbacher.
\newblock The effect of brochure use versus therapist teaching on patients performing therapeutic exercise and on changes in impairment status.
\newblock {\em Physical Therapy}, 76(10):1082--1088, Oct. 1996.

\bibitem{bib3}
M.~C. Fu, A.~M. Samuel, P.~K. Sculco, C.~H. MacLean, D.~E. Padgett, and A.~S. McLawhorn.
\newblock Discharge to inpatient facilities after total hip arthroplasty is associated with increased postdischarge morbidity.
\newblock {\em The Journal of Arthroplasty}, 32(9S):S144--S149.e1, Sept. 2017.

\bibitem{bib25}
S.~Greenland, S.~J. Senn, K.~J. Rothman, J.~B. Carlin, C.~Poole, S.~N. Goodman, et~al.
\newblock Statistical tests, p values, confidence intervals, and power: A guide to misinterpretations.
\newblock {\em European Journal of Epidemiology}, 31(4):337--350, Apr. 2016.

\bibitem{bib28}
N.~Jirasakulsuk, P.~Saengpromma, and S.~Khruakhorn.
\newblock Real-time telerehabilitation in older adults with musculoskeletal conditions: Systematic review and meta-analysis.
\newblock {\em JMIR Rehabilitation and Assistive Technologies}, 9(3):e36028, Sept. 2022.

\bibitem{bib10}
M.~Kafri, M.~J. Myslinski, V.~K. Gade, and J.~E. Deutsch.
\newblock Energy expenditure and exercise intensity of interactive video gaming in individuals poststroke.
\newblock {\em Neurorehabilitation and Neural Repair}, 28(1):56--65, Jan. 2014.

\bibitem{bib34}
C.~Killingback, F.~Tsofliou, and C.~Clark.
\newblock Older people’s adherence to community-based group exercise programmes: A multiple-case study.
\newblock {\em BMC Public Health}, 17, Jan. 2017.

\bibitem{bib29}
V.~Ko, J.~Naylor, I.~Harris, J.~Crosbie, A.~Yeo, and R.~Mittal.
\newblock One-to-one therapy is not superior to group or home-based therapy after total knee arthroplasty: A randomized, superiority trial.
\newblock {\em The Journal of Bone and Joint Surgery}, 95(21):1942--1949, Nov. 2013.

\bibitem{bib15}
L.~M. Koonin, B.~Hoots, C.~A. Tsang, Z.~Leroy, K.~Farris, T.~Jolly, et~al.
\newblock Trends in the use of telehealth during the emergence of the covid-19 pandemic - united states, january-march 2020.
\newblock {\em MMWR Morbidity and Mortality Weekly Report}, 69(43):1595--1599, Oct. 2020.

\bibitem{bib17}
D.~Levac, D.~Espy, E.~Fox, S.~Pradhan, and J.~E. Deutsch.
\newblock “kinect-ing” with clinicians: A knowledge translation resource to support decision making about video game use in rehabilitation.
\newblock {\em Physical Therapy}, 95(3):426--440, Mar. 2015.

\bibitem{bib30}
M.~O'Keeffe, A.~Hayes, K.~McCreesh, H.~Purtill, and K.~O'Sullivan.
\newblock Are group-based and individual physiotherapy exercise programmes equally effective for musculoskeletal conditions? a systematic review and meta-analysis.
\newblock {\em British Journal of Sports Medicine}, 51(2):126--132, Jan. 2017.

\bibitem{bib11}
M.~A. O'Reilly, P.~Slevin, T.~Ward, and B.~Caulfield.
\newblock A wearable sensor-based exercise biofeedback system: Mixed methods evaluation of formulift.
\newblock {\em JMIR MHealth UHealth}, 6(1):e33, Jan. 2018.

\bibitem{bib7}
C.~Palazzo, E.~Klinger, V.~Dorner, A.~Kadri, O.~Thierry, Y.~Boumenir, et~al.
\newblock Barriers to home-based exercise program adherence with chronic low back pain: Patient expectations regarding new technologies.
\newblock {\em Annals of Physical and Rehabilitation Medicine}, 59(2):107--113, Apr. 2016.

\bibitem{bib36}
M.~Pantell, D.~Rehkopf, D.~Jutte, S.~L. Syme, J.~Balmes, and N.~Adler.
\newblock Social isolation: A predictor of mortality comparable to traditional clinical risk factors.
\newblock {\em American Journal of Public Health}, 103(11):2056--2062, Nov. 2013.

\bibitem{bib6}
M.~F. Pisters, C.~Veenhof, F.~G. Schellevis, J.~W.~R. Twisk, J.~Dekker, and D.~H. De~Bakker.
\newblock Exercise adherence improving long-term patient outcome in patients with osteoarthritis of the hip and/or knee.
\newblock {\em Arthritis Care \& Research}, 62(8):1087--1094, Aug. 2010.

\bibitem{bib35}
G.~Richardson, N.~Hawkins, C.~J. McCarthy, P.~M. Mills, R.~Pullen, C.~Roberts, et~al.
\newblock Cost-effectiveness of a supplementary class-based exercise program in the treatment of knee osteoarthritis.
\newblock {\em International Journal of Technology Assessment in Health Care}, 22(1):84--89, 2006.

\bibitem{bib23}
K.~J. Rothman.
\newblock No adjustments are needed for multiple comparisons.
\newblock {\em Epidemiology}, 1(1):43--46, Jan. 1990.

\bibitem{bib16}
T.~G. Russell.
\newblock Telehealth for musculoskeletal physiotherapy.
\newblock {\em Musculoskeletal Science and Practice}, 48:102193, Aug. 2020.

\bibitem{bib18}
G.~Saposnik, M.~Levin, and O.~R. C. S.~W. Group.
\newblock Virtual reality in stroke rehabilitation: A meta-analysis and implications for clinicians.
\newblock {\em Stroke}, 42(5):1380--1386, May 2011.

\bibitem{bib24}
D.~J. Saville.
\newblock Multiple comparison procedures: The practical solution.
\newblock {\em The American Statistician}, 44(2):174--180, May 1990.

\bibitem{bib1}
M.~Sloan, A.~Premkumar, and N.~P. Sheth.
\newblock Projected volume of primary total joint arthroplasty in the {U.S.}, 2014 to 2030.
\newblock {\em The Journal of Bone and Joint Surgery. Am.}, 100(17):1455--1460, Sept. 2018.

\bibitem{bib12}
L.~Suso-Martí, R.~La~Touche, A.~Herranz-Gómez, S.~Angulo-Díaz-Parreño, A.~Paris-Alemany, and F.~Cuenca-Martínez.
\newblock Effectiveness of telerehabilitation in physical therapist practice: An umbrella and mapping review with meta-meta-analysis.
\newblock {\em Physical Therapy}, 101(5):pzab075, May 2021.

\bibitem{bib37}
H.~O. Taylor, R.~J. Taylor, A.~W. Nguyen, and L.~Chatters.
\newblock Social isolation, depression, and psychological distress among older adults.
\newblock {\em Journal of Aging and Health}, 30(2):229--246, Feb. 2018.

\bibitem{bib31}
E.~Toomey, L.~Currie-Murphy, J.~Matthews, and D.~A. Hurley.
\newblock The effectiveness of physiotherapist-delivered group education and exercise interventions to promote self-management for people with osteoarthritis and chronic low back pain: A rapid review part i.
\newblock {\em Manuel Therapy}, 20(2):265--286, Apr. 2015.

\bibitem{bib13}
M.~P. Tsang, G.~C.~W. Man, H.~Xin, Y.~C. Chong, M.~T.~Y. Ong, and P.~S.~H. Yung.
\newblock The effectiveness of telerehabilitation in patients after total knee replacement: A systematic review and meta-analysis of randomized controlled trials.
\newblock {\em Journal of Telemedicine and Telecare}, 1357633X221097469:N/A, May 2022.

\bibitem{bib8}
D.~E.~R. Warburton, S.~S.~D. Bredin, L.~T.~L. Horita, D.~Zbogar, J.~M. Scott, B.~T.~A. Esch, et~al.
\newblock The health benefits of interactive video game exercise.
\newblock {\em Applied Physiology, Nutrition, and Metabolism}, 32(4):655--663, Aug. 2007.

\bibitem{bib19}
P.~N. Wilson, N.~Foreman, and D.~Stanton.
\newblock Virtual reality, disability and rehabilitation.
\newblock {\em Disability and Rehabilitation}, 19(6):213--220, Jan. 1997.

\bibitem{bib20}
J.~Wosik, M.~Fudim, B.~Cameron, Z.~F. Gellad, A.~Cho, D.~Phinney, et~al.
\newblock Telehealth transformation: Covid-19 and the rise of virtual care.
\newblock {\em Journal of the American Medical Informatics Association}, 27(6):957--962, June 2020.

\end{thebibliography}
\end{document}